\documentclass{article}

\usepackage{ifthen}
\newboolean{neurips_version}
\setboolean{neurips_version}{true}
\ifthenelse{\boolean{neurips_version}}
{\PassOptionsToPackage{numbers, compress}{natbib}
 \usepackage[preprint]{neurips_2022}
}
{
\usepackage{PRIMEarxiv}
 \usepackage[numbers]{natbib}
}

\usepackage[utf8]{inputenc} 
\usepackage[T1]{fontenc}    
\usepackage{hyperref}       
\usepackage{url}            
\usepackage{booktabs}       
\usepackage{amsfonts,amsmath,amsthm,amssymb}       
\usepackage{nicefrac}       
\usepackage{microtype}      
\usepackage{xcolor}         
\usepackage[disable]{todonotes}
\usepackage{enumitem}
\usepackage{caption}

\usepackage{mathtools} 
\usepackage{tikz}
\usetikzlibrary{calc}
\usepackage{pgfplots}
\usepackage{graphicx}

\usepackage{amsmath}
\usepackage{algorithm, algpseudocode}
\usepackage{amsfonts}  

\usepackage[english]{babel}
\usepackage{amsthm}

\usepackage{subfig}
\usepackage[export]{adjustbox}

\usepackage{hhline}
\usepackage{makecell, caption, booktabs}
\usepackage{siunitx}

\usepackage{multirow}

\usepackage{amsmath, nccmath}
\usepackage{bigstrut}

\usepackage{booktabs}

\usepackage{lipsum}
\graphicspath{{media/}}     

\usepackage{CJKutf8}
\usepackage[framed,numbered,autolinebreaks,useliterate]{mcode}

\title{ CryptoUNets: Applying Convolutional Networks to Encrypted Data for Biomedical Image Segmentation  }

\author{ \href{https://orcid.org/0000-0003-0378-0607}{\includegraphics[scale=0.06]{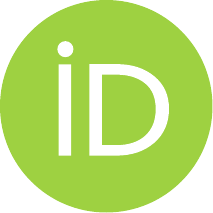}\hspace{1mm}John Chiang} \\                             
                                      \\
	\texttt{john.chiang.smith@gmail.com} 
}

\date{}

\theoremstyle{remark}

\renewcommand{\epsilon}{\varepsilon}

\makeatletter
\def\namedlabel#1#2{\begingroup
    #2%
    \def\@currentlabel{#2}%
    \phantomsection\label{#1}\endgroup
}
\makeatother

\algnewcommand{\LeftComment}[1]{\Statex \(\triangleright\) #1}
\algnewcommand{\LineCommentStep}[1]{\Statex \textbf{[Step #1]:} }
\makeatletter
\newlength{\trianglerightwidth}
\algnewcommand{\LineComment}[1]{\Statex \hskip\ALG@thistlm $\triangleright$ #1}
\algnewcommand{\LineCommentCont}[1]{\Statex \hskip\ALG@thistlm%
  \parbox[t]{\dimexpr\linewidth-\ALG@thistlm}{\hangindent=\trianglerightwidth \hangafter=1 \strut$\triangleright$ #1\strut}}
\algnewcommand{\LeftLineCommentCont}[1]{\Statex \hskip\ALG@thistlm%
  \parbox[t]{\dimexpr\linewidth-\ALG@thistlm}{\leftskip=\algorithmicindent \hangindent=\trianglerightwidth \hangafter=1 \strut$\triangleright$ #1\strut}}

\newcommand{\mysplit}[1]{%
  \begin{tabular}{@{}c@{}}
    #1
  \end{tabular}
  }
  
\begin{document}

\maketitle

\begin{abstract}%

In this manuscript, we demonstrate the feasibility of a privacy-preserving U-Net deep learning inference framework, namely, homomorphic encryption-based U-Net inference. That is, U-Net inference can be performed solely using homomorphic encryption techniques. To our knowledge, this is the first work to achieve support perform implement enable U-Net inference entirely based on homomorphic encryption ?.

The primary technical challenge lies in data encoding. To address this, we employ a flexible encoding scheme, termed Double Volley Revolver, which enables effective support for skip connections and upsampling operations within the U-Net architecture.

We adopt a tailored HE-friendly U-Net design incorporating square activation functions, mean pooling layers, and transposed convolution layers (implemented as ConvTranspose2d in PyTorch) with a kernel size of 2 and stride of 2. After training the model in plaintext, we deploy the resulting parameters using the HEAAN homomorphic encryption library to perform encrypted U-Net inference.

The complete, runnable C++ code to implement our  work can be found at: \href{https://github.com/petitioner/HE.CryptoUNets}{$\texttt{https://github.com/petitioner/HE.CryptoUNets}$}. 

\end{abstract}

\listoftodos

\section{Introduction}

\subsection{Background}

Deep neural networks (DNNs) have emerged as powerful and versatile tools across a broad range of domains, including speech recognition, natural language processing, and computer vision. Their deployment typically involves two key phases: training and inference. In the training phase, a suitable dataset is selected, a network architecture is designed, and the model is optimized by adjusting its parameters over multiple epochs. This process often requires significant computational resources and time, potentially spanning several days. Once trained, the network enters the inference phase, where it is expected to generate predictions for unseen inputs efficiently.

Despite the utility of DNNs, many real-world applications involve sensitive data that cannot be openly shared. For instance, credit card transaction records are proprietary to financial institutions, and healthcare datasets—such as patient histories or diagnostic imagery—are restricted to clinical entities due to legal and ethical constraints. Regulations like the General Data Protection Regulation (GDPR) in the European Union further limit data access and sharing. In many cases, data custodians lack the technical expertise to train DNNs themselves, and privacy concerns prevent them from outsourcing the task to external service providers.

Fully Homomorphic Encryption (FHE) presents a promising solution for reconciling the competing demands of data privacy and machine learning. Although early perceptions deemed FHE computationally impractical, substantial progress over the last decade—both in theoretical advances and practical implementations—has enabled research prototypes to apply FHE in a variety of domains. While FHE-based training of deep neural networks remains computationally intensive and largely impractical at scale, the inference phase under FHE is increasingly considered viable and has attracted growing attention.

\subsection{Related Work}

Privacy-preserving machine learning has been an active research area for nearly two decades. However, the application of homomorphic encryption specifically to deep neural networks remains relatively underexplored. To date, one of the most notable efforts in this direction is CryptoNets by Gilad-Bachrach et al., which introduced a carefully constructed neural network capable of performing inference directly on encrypted inputs using non-interactive homomorphic encryption. Their model achieved 99\% accuracy on the MNIST handwritten digit classification task and an amortized throughput of approximately 60,000 predictions per hour.

More recently, fully homomorphic encryption has also been adopted in other privacy-preserving tasks, such as secure face matching and encrypted $k$-nearest neighbor search. Nevertheless, these works are largely limited to inference.

A more extensive body of research has explored hybrid protocols that combine homomorphic encryption with interactive secure computation. Early examples include the works of Barni et al. and Orlandi et al., which employed additively homomorphic encryption in conjunction with interactive protocols to achieve inference on small-scale networks in around 10 seconds. Subsequent frameworks—such as SecureML by Mohassel and Zhang, MiniONN by Liu et al., Chameleon by Riazi et al., and GAZELLE by Juvekar et al.—have significantly improved both accuracy and latency. For instance, GAZELLE performs inference on MNIST in 30 ms and on CIFAR-10 in approximately 13 seconds.

Despite these advances, most of these studies focus exclusively on the inference stage and do not address the challenge of privacy-preserving training. Only a handful of works have explored this direction~\cite{nandakumar2019towards, chiang2023privacy3layer}, reflecting a broader skepticism about the practicality of training neural networks under FHE due to its perceived inefficiency.

In this work, we take a further step toward expanding the applicability of homomorphic encryption in deep learning by demonstrating the feasibility of executing U-Net inference entirely within the encrypted domain.

\subsection{Contributions}

This paper proposes a method for performing U-Net inference on encrypted data using Fully Homomorphic Encryption (FHE). Our work focuses exclusively on the inference phase, under the assumption that the model—specifically, a U-Net architecture—has been pre-trained offline on plaintext data and is already available on the cloud.

Although privacy-preserving model training is a critical and challenging problem, especially in domains where data confidentiality is paramount, our study does not attempt to address this issue. Instead, we aim to demonstrate that FHE-based inference can be both practical and accurate, even when applied to complex models like U-Net.

A frequent criticism of homomorphic encryption is its high computational overhead, which has historically led to the belief that it is impractical for most machine learning applications. However, by integrating techniques from cryptography, deep learning, and systems engineering, we show that it is possible to construct an FHE-based pipeline that supports real-world inference scenarios with acceptable performance. In this context, our work can be seen as a continuation of the CryptoNets paradigm, extending its applicability to more sophisticated neural architectures and more demanding vision tasks.

\section{Preliminaries}

\subsection{Fully Homomorphic Encryption}

Homomorphic Encryption (HE) encompasses a class of cryptographic primitives that enable computation over ciphertexts, such that the decrypted result of the homomorphic operation corresponds to the outcome of the same operation performed on the plaintexts. A scheme is termed \emph{fully} homomorphic encryption (FHE) if it supports both additive and multiplicative operations, thereby achieving Turing-completeness over encrypted inputs. Gentry's seminal construction in 2009~\cite{gentry2009fully} introduced the first plausible FHE scheme by employing ideal lattices, bootstrapping, and a novel approach to managing noise growth.

Subsequent advancements have significantly improved the asymptotic and concrete efficiency of FHE. Notably, Brakerski, Gentry, and Vaikuntanathan~\cite{brakerski2014leveled} proposed the BGV leveled FHE scheme, which eliminates bootstrapping in exchange for circuit depth restrictions. The introduction of the \emph{SIMD (Single Instruction, Multiple Data)} paradigm by Smart and Vercauteren~\cite{SmartandVercauteren_SIMD}, leveraging the Chinese Remainder Theorem (CRT) decomposition of plaintext polynomials, enables the packing of multiple plaintext slots into a single ciphertext, thus facilitating parallelized computation.

A critical development enabling approximate arithmetic for real-number computations—especially relevant in privacy-preserving machine learning—is the \emph{rescaling} operation introduced in the CKKS scheme~\cite{cheon2017homomorphic}. Rescaling effectively manages the ciphertext modulus and controls the scale factor to bound precision loss and mitigate noise accumulation.

Modern approximate FHE libraries, such as \texttt{HEAAN}, support a rich set of homomorphic operations:
\begin{itemize}
    \item $\texttt{Enc}$: Encryption of a plaintext vector into a ciphertext;
    \item $\texttt{Dec}$: Decryption of a ciphertext into its corresponding plaintext;
    \item $\texttt{Add}$, $\texttt{Mult}$: Ciphertext-ciphertext addition and multiplication;
    \item $\texttt{cMult}$: Ciphertext-plaintext multiplication with a constant vector;
    \item $\texttt{ReScale}$: Scale reduction operation following multiplication to maintain modulus alignment;
    \item $\texttt{Rot}$: Encrypted vector rotation, often used for index permutation;
    \item $\texttt{bootstrap}$: Noise-refreshing procedure via homomorphic decryption and reencryption.
\end{itemize}

These primitives collectively enable the implementation of expressive encrypted computations, serving as the foundation for privacy-preserving machine learning, encrypted database query processing, and secure multiparty computation.

\subsubsection{Data Encoding via Slot-Packing}

To maximize ciphertext slot utilization in homomorphic matrix processing, Kim et al.~\cite{IDASH2018Andrey} proposed a slot-packing scheme that enables efficient plaintext-to-ciphertext encoding for tabular datasets. Given a row-major matrix $Z \in \mathbb{R}^{n \times d}$, the matrix is first linearized into a one-dimensional vector $V \in \mathbb{R}^{n \cdot d}$ using row-wise flattening. The encrypted matrix is then represented as $Z = \texttt{Enc}(V)$.

Based on this encoding, two core homomorphic transformations are performed using rotation operations:
\begin{itemize}
    \item \textbf{Full Row Rotation}: Cyclically shifts entire row segments across the vector;
    \item \textbf{Partial Column Rotation}: Simulates per-column operations by offsetting specific column-aligned entries across adjacent rows.
\end{itemize}

These transformations enable encrypted matrix manipulation in the ciphertext domain, producing the following permuted encodings:

\begin{equation*}
 \begin{aligned}
 Z &= 
\left[ \begin{array}{cccc}
 x_{10}  &   x_{11}  &  \ldots  &  x_{1d}  \\
 x_{20}  &   x_{21}  &  \ldots  &  x_{2d}  \\
 \vdots         &   \vdots         &  \ddots  &  \vdots         \\
 x_{n0}  &  x_{n1}   &  \ldots  &  x_{nd}  \\
 \end{array}
 \right], 
& Z^{'}  = \texttt{Enc}
\left[ \begin{array}{cccc}
 x_{20}  &   x_{21}  &  \ldots  &  x_{2d}  \\
 \vdots         &   \vdots         &  \ddots  &  \vdots         \\
 x_{n0}  &   x_{n1}  &  \ldots  &  x_{nd}  \\
 x_{10}  &  x_{11}   &  \ldots  &  x_{1d}  \\
 \end{array}
 \right],   \\
 Z^{''}  &= \texttt{Enc}
\left[ \begin{array}{cccc}
 x_{11}  &  \ldots  &  x_{1d}  &   x_{20}  \\
 x_{21}  &  \ldots  &  x_{2d}  &   x_{30}  \\
 \vdots         &   \vdots         &  \ddots  &  \vdots         \\
 x_{n1}  &  \ldots  &  x_{nd}  &   x_{10} \\
 \end{array}
 \right], 
 & Z^{'''} = \texttt{Enc}
\left[ \begin{array}{cccc}
 x_{11}  &  \ldots  &  x_{1d}  &   x_{10}  \\
 x_{21}  &  \ldots  &  x_{2d}  &   x_{20}  \\
 \vdots         &   \vdots         &  \ddots  &  \vdots         \\
 x_{n1}  &  \ldots  &  x_{nd}  &   x_{n0} \\
 \end{array}
 \right]  .
 \end{aligned}
\end{equation*}

The transformation from $Z$ to $Z^{'''}$—which achieves full column rotation—can be realized via a composition of two $\texttt{Rot}$ operations, two $\texttt{cMult}$ operations for masking, and one $\texttt{Add}$ to combine intermediate ciphertexts.

Extensions to this encoding framework~\cite{han2018efficient, chiang2022novel} have incorporated utility functions such as $\texttt{SumRowVec}$ and $\texttt{SumColVec}$, which enable ciphertext-domain summation across rows and columns, respectively. These primitives are instrumental in implementing homomorphic reductions, gradient aggregation, and other linear algebra operations central to encrypted machine learning pipelines.

\subsection{Fully Convolutional Networks}
Convolutional Neural Networks (CNNs) constitute a prominent class of deep learning models, architecturally inspired by the hierarchical structure of the biological visual cortex. Owing to their ability to capture spatial hierarchies in data through local receptive fields and weight sharing, CNNs have achieved state-of-the-art performance across a wide range of computer vision tasks, particularly in image classification. Their biologically inspired design sets them apart as one of the few neural architectures that closely emulate the functional organization of the human visual system.

Long et al.~\cite{long2014fully} proposed a more elegant architecture, known as the fully convolutional network (FCN), to address the challenge of dense prediction in semantic segmentation.
The main idea in~\cite{long2014fully} is to augment the conventional contracting network with a sequence of layers in which pooling operations are replaced by upsampling operators. These additional layers progressively restore spatial resolution in the output. To preserve localization accuracy, high-resolution features from the contracting path are combined with the upsampled activations. A subsequent convolutional layer is then capable of learning to generate more refined predictions based on this integrated information.

\textbf{Fully Convolutional Architecture:}
Traditional convolutional neural networks (CNNs) employ fully connected layers in the final stages for classification, which constrains input images to a fixed size. FCNs eliminate these fully connected layers and instead utilize convolutional layers throughout the entire network, enabling the model to accept arbitrary-sized inputs and produce correspondingly sized dense predictions. This fully convolutional design allows the network to perform pixel-wise predictions on entire images without requiring cropping or patch-based processing.

\textbf{Upsampling and Deconvolution:}
Since convolutions and pooling layers reduce the spatial resolution of feature maps, FCNs incorporate deconvolutional (or transposed convolutional) layers to upsample the low-resolution outputs back to the original input dimensions. The parameters of these deconvolution layers are learned via backpropagation, allowing the model to reconstruct high-resolution outputs that align with the input image size.

\textbf{Skip Connections:}
To combine semantic information from deep layers with fine-grained details from shallow layers, FCNs introduce skip connections. Specifically, high-resolution feature maps from earlier layers in the contracting path are fused with the corresponding upsampled outputs. This fusion enhances spatial precision in the final predictions. The skip connection mechanism is conceptually similar to that later used in the U-Net architecture.

\textbf{End-to-End Training:}
FCNs enable end-to-end learning from raw input images to dense segmentation maps. Unlike traditional approaches that require handcrafted pre-processing and post-processing pipelines, FCNs streamline the training process and improve efficiency by jointly optimizing the entire network in a single training loop.

Ronneberger et al.~\cite{ronneberger2015u} proposed the U-Net architecture, a modification and extension of the Fully Convolutional Network (FCN), designed to achieve precise and accurate segmentation results even with limited training data.

\subsection{Double Volley Revolver}

Unlike other efficient yet intricate encoding schemes~\citep{kim2018matrix}, \texttt{Volley Revolver}~\citep{chiang2022novel} is a lightweight and flexible matrix encoding technique specifically designed for privacy-preserving machine learning. Its core idea, in its simplest form, involves encrypting the transpose of one of the matrices involved in a matrix multiplication, thereby enabling efficient homomorphic computation.

Encoding schemes play a pivotal role in enabling privacy-preserving training of convolutional neural networks (CNNs). As demonstrated in~\citep{chiang2022novel}, \texttt{Volley Revolver} can be effectively utilized for implementing homomorphic CNN training. Despite its simplicity, this encoding strategy allows fine-grained control over data movement within ciphertexts, facilitating practical encrypted computation.

Importantly, it is not mandatory to always transpose the second matrix. In practice, either of the two matrices can be transposed prior to encryption. For example, transposing the first matrix and adapting the multiplication accordingly leads to an algorithm similar to Algorithm~2 in~\citep{chiang2022novel}.

Furthermore, when the matrices involved are too large to be encapsulated in a single ciphertext, we extend the original scheme to a batched version, termed \texttt{Double Volley Revolver}. In this setting, each matrix is partitioned horizontally into multiple ciphertexts, forming two teams $A$ and $B$. The outer loop of the algorithm iterates over all ciphertext pairs from these two sets, while the inner loop performs submatrix multiplication between each pair $(A_{[i]}, B_{[j]})$ using the original \texttt{Volley Revolver} algorithm.

\subsubsection{Horizontal Partitioning}

\begin{figure}[htp]
\centering
\includegraphics[scale=0.6]{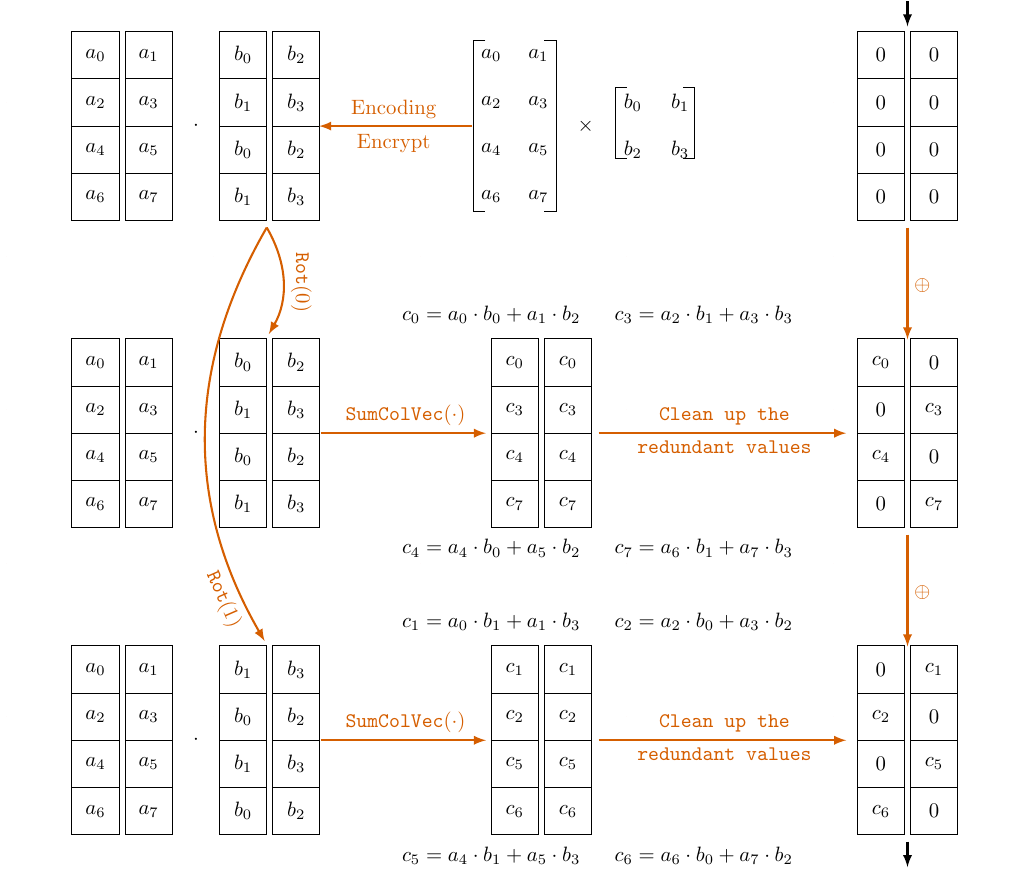}
\caption{The matrix multiplication algorithm of \texttt{Volley Revolver} applied to a $4 \times 2$ matrix $A$ and a $2 \times 2$ matrix $B$.}
\label{Matrix Multiplication}
\end{figure}

Figure~\ref{Matrix Multiplication} illustrates a simple example of the multiplication process used in this encoding scheme.

\section{Technical Details}
%

\subsection{U-Net Convolutional Networks}

U-Net is a specialized convolutional neural network (CNN) architecture primarily designed for semantic segmentation tasks, particularly in the field of medical image analysis. It is well-suited for tasks such as organ segmentation, tumor detection, and other pixel-level image classification problems, where precise localization and boundary delineation are crucial. The network's architecture is characterized by an encoder-decoder structure, where the encoder captures high-level features through successive downsampling, and the decoder reconstructs the spatial resolution through upsampling, ultimately providing pixel-wise predictions.

One of the key strengths of U-Net is its ability to handle limited datasets, making it highly effective in medical imaging where annotated data is often scarce. This is achieved through the use of skip connections that enable the network to retain fine-grained spatial information while learning hierarchical features. U-Net has demonstrated state-of-the-art performance in various medical image segmentation benchmarks, establishing itself as a cornerstone in medical imaging research.

The architecture of U-Net relies on four primary techniques to achieve its impressive performance: convolution, pooling, upsampling, and skip connections. Each of these components plays a critical role in feature extraction, resolution reduction, feature map restoration, and the retention of spatial details, respectively. These techniques are detailed in the following sections.

\subsubsection{Convolution}

Convolution is the fundamental operation for feature extraction in convolutional neural networks. It involves applying small-sized kernels (e.g., $3 \times 3$) that slide over the input image to capture local spatial patterns such as edges, textures, and corners. Through the stacking of multiple convolutional layers, the network can progressively extract increasingly abstract representations—ranging from low-level visual cues to higher-level semantic structures.

In U-Net, each convolutional block typically consists of two successive $3 \times 3$ convolutional layers, each followed by a Rectified Linear Unit (ReLU) activation function. This configuration preserves fine-grained spatial details while simultaneously enriching the representation capacity of the network.

\subsubsection{Pooling}

Pooling serves to reduce the spatial dimensions of feature maps, enabling the network to abstract higher-level representations while reducing computational complexity. Max pooling, which selects the maximum value within a local region (typically $2 \times 2$), is commonly used due to its ability to retain the most salient features.

This operation offers two key advantages: (1) it significantly decreases the number of parameters and computational overhead, and (2) it increases the receptive field, allowing the model to capture more global contextual information. In U-Net, downsampling is implemented using $2 \times 2$ max pooling operations, which halve the spatial resolution at each stage.

\subsubsection{Upsampling}

Upsampling is employed to restore the spatial resolution of feature maps that have been compressed during the encoding phase. Basic methods such as nearest-neighbor interpolation or bilinear interpolation can be used for this purpose; however, these approaches are fixed and non-learnable.

A more effective and widely adopted method in U-Net is transposed convolution (also known as deconvolution), which learns a set of weights to perform upsampling in a data-driven manner. This learnable upsampling mechanism enables the network to recover finer structural details and improve localization accuracy.

\subsubsection{Skip Connections}

Skip connections are a core architectural component of U-Net, designed to mitigate the loss of spatial detail caused by repeated convolution and pooling operations. As the network goes deeper, spatial resolution is progressively reduced, making it challenging to accurately localize fine-grained structures such as object boundaries.

To address this, U-Net introduces skip connections that directly concatenate the feature maps from the encoder (at a given resolution level) with the corresponding decoder feature maps after upsampling. This operation, performed along the channel dimension, allows the decoder to leverage both high-level semantic information and low-level spatial details.

This design greatly enhances the network's ability to perform precise pixel-level predictions, and is widely regarded as one of the most influential innovations introduced by the U-Net architecture.

\subsection{Homomorphic U-Net Inference} 

\subsubsection{Polynomial Approximation}

Polynomial approximation techniques are central to many applications in numerical analysis, particularly when seeking efficient representations of nonlinear functions. Classical approaches, such as \emph{Taylor expansion} and \emph{Lagrange interpolation}, offer high accuracy in localized regions around a specific expansion point. However, their utility diminishes rapidly outside the neighborhood of the expansion point, with the approximation error increasing exponentially as the input moves further from this vicinity.

In contrast, the \emph{least squares approximation} method minimizes the overall approximation error across a global domain, yielding a more reliable and stable representation over broader intervals. The least squares framework optimizes the coefficients of a polynomial by minimizing the sum of squared errors between the approximated polynomial and the target function. This method's robustness and generality have made it a popular choice in practical applications, as evidenced by its integration in software packages such as Python's \texttt{polyfit(·)} and MATLAB's \texttt{polyfit}. These built-in functions efficiently implement least squares polynomial fitting for a wide range of non-polynomial functions, offering significant flexibility in modeling real-world data.

Another widely adopted technique is the \emph{minimax approximation}, which minimizes the maximum error across the entire approximation interval. This approach ensures a uniform approximation quality, making it particularly valuable in applications where tight bounds on worst-case errors are critical. The minimax method has found broad application in the context of function approximation where control over the peak error is essential.

Recent advancements have extended polynomial approximation techniques to large intervals, often incorporating domain-specific optimizations. Cheon et al.~\cite{cheon2022efficient} introduced the concept of \emph{domain extension polynomials}, a strategy that facilitates the iterative extension of the approximation domain. This method proves particularly useful for approximating functions that exhibit behavior analogous to sigmoid functions, enabling the efficient approximation of such functions over significantly broader intervals. This technique has found specific utility in \emph{homomorphic evaluation}, where accurate function approximations are required within encrypted computation frameworks. Building on this work, we adopt their methodology to approximate the sigmoid function over the interval $[-64, 64]$, leveraging the extended domain to achieve a balance between computational efficiency and approximation fidelity.

These techniques collectively form a powerful toolkit for approximating complex nonlinear functions, ensuring high-quality results across both narrow and broad domains, and providing a basis for efficient encrypted function evaluation in privacy-preserving applications.

\subsubsection{Homomorphic Evaluation }
Here is a description of the network used for training:
\begin{lstlisting}
Model Summary:
UNet(
  (conv11): Conv2d(1, 5, kernel_size=(3, 3), stride=(1, 1), padding=(1, 1))
  (act11): MyReLU()
  (conv12): Conv2d(5, 5, kernel_size=(3, 3), stride=(1, 1), padding=(1, 1))
  (act12): MyReLU()
  (pool1): AvgPool2d(kernel_size=2, stride=2, padding=0)
  (conv21): Conv2d(5, 10, kernel_size=(3, 3), stride=(1, 1), padding=(1, 1))
  (act21): MyReLU()
  (conv22): Conv2d(10, 10, kernel_size=(3, 3), stride=(1, 1), padding=(1, 1))
  (act22): MyReLU()
  (pool2): AvgPool2d(kernel_size=2, stride=2, padding=0)
  (conv51): Conv2d(10, 20, kernel_size=(3, 3), stride=(1, 1), padding=(1, 1))
  (act51): MyReLU()
  (conv52): Conv2d(20, 20, kernel_size=(3, 3), stride=(1, 1), padding=(1, 1))
  (act52): MyReLU()
  (convtran1): ConvTranspose2d(20, 10, kernel_size=(2, 2), stride=(2, 2))
  (conv61): Conv2d(20, 10, kernel_size=(3, 3), stride=(1, 1), padding=(1, 1))
  (act61): MyReLU()
  (conv62): Conv2d(10, 10, kernel_size=(3, 3), stride=(1, 1), padding=(1, 1))
  (act62): MyReLU()
  (convtran2): ConvTranspose2d(10, 5, kernel_size=(2, 2), stride=(2, 2))
  (conv71): Conv2d(10, 5, kernel_size=(3, 3), stride=(1, 1), padding=(1, 1))
  (act71): MyReLU()
  (conv72): Conv2d(5, 5, kernel_size=(3, 3), stride=(1, 1), padding=(1, 1))
  (act72): MyReLU()
  (out): Conv2d(5, 1, kernel_size=(1, 1), stride=(1, 1))
)
\end{lstlisting}

\subsection{Loss Function}

The loss function plays a crucial role in guiding the network to learn accurate semantic segmentation. In the context of U-Net and similar architectures, the most commonly employed loss function is the \textit{binary cross-entropy loss}, which treats segmentation as a pixel-wise binary classification problem. Each pixel is assigned a label (e.g., foreground or background), and the loss evaluates the discrepancy between the predicted probability and the ground truth label at every pixel.

Binary cross-entropy (BCE) loss originates from the concept of cross-entropy in information theory, introduced by Shannon (1948), and is formally defined as the negative log-likelihood under a Bernoulli distribution. It is particularly suitable for settings where the model outputs a probability value in the range $[0, 1]$ for each pixel, indicating the likelihood of belonging to the positive class.

However, BCE may become suboptimal in scenarios involving class imbalance—such as when the foreground region occupies only a small portion of the image. To address this, alternative loss functions such as the \textit{Dice loss} and the \textit{Intersection over Union (IoU) loss} have been widely adopted. Dice loss, in particular, is effective at handling highly imbalanced segmentation tasks, as it directly optimizes for overlap between predicted and ground-truth masks.

In the original U-Net architecture, the authors employed standard pixel-wise binary cross-entropy loss. In subsequent variants and extensions of U-Net, Dice-based loss functions have been introduced, often leading to improved performance, especially in medical image segmentation tasks where foreground objects tend to be small and sparse. Despite the potential benefits of these alternatives, we follow the original U-Net implementation and adopt binary cross-entropy loss due to its simplicity and ease of integration. Specifically, we utilize the \texttt{nn.BCEWithLogitsLoss()} function provided by PyTorch, which combines a sigmoid layer and the BCE loss in a numerically stable implementation.

\subsubsection{Computational Complexity}

Experiments for encrypted-domain processing were conducted on a dual-socket Intel Xeon E5-2698 v3 server, based on the Haswell microarchitecture. This server configuration includes 16 cores per socket running at 2.30 GHz, with a total of 250~GB of main memory. The software environment utilized GCC version 7.2.1 for compilation, with arithmetic operations supported by NTL version 10.5.0 and GMP version 6.0.

\textbf{Future improvements:} The preliminary results presented herein validate the feasibility of performing Stochastic Gradient Descent (SGD) training within the encrypted domain. However, this work represents a nascent phase, and several avenues for optimization remain unexplored. Notably, our current approach only addresses basic batching techniques for input data, while still allocating individual ciphertexts for each weight parameter. Efficient strategies for ciphertext packing and optimized batching, which could substantially enhance computational efficiency, are currently under investigation.

\section{Empirical Results}
The C++ source code to implement the experiments in this section  is openly available at: \href{https://github.com/petitioner/HE.CryptoUNets}{$\texttt{https://github.com/petitioner/HE.CryptoUNets}$} .

\subsection{Dataset}

To assess the performance of our U-Net-based segmentation method, we utilize the ISBI 2012 EM Segmentation Challenge dataset. This dataset comprises a series of electron microscopy (EM) images capturing neuronal structures from the Drosophila larval ventral nerve cord. The primary objective of the challenge is to achieve precise segmentation of cell membranes within these grayscale images, with an emphasis on delineating fine structural boundaries and mitigating potential noise within the dataset.

\subsection{Parameters}

The parameters of $\texttt{HEAAN}$ utilized in our experiments are as follows: $logN = 16$, $logQ = 990$, $logp = 45$, and $slots = 32768$, corresponding to a security level of $\lambda = 128$. Further details regarding these parameter choices can be found in \cite{IDASH2018Andrey}. Notably, bootstrapping was not applied to refresh the weight ciphertexts. Each input image incurs an approximate runtime of 11 minutes, with a peak memory usage of approximately 18 GB.

\subsection{Performance}
\subsection{Timing analysis}
\subsection{Description of the Network}
\subsection{Message sizes}

\section{Conclusion}

In this work, we implemented privacy-preserving U-Net inference solely based on homomorphic encryption techniques by employing a flexible data encoding scheme.

\bibliography{HE.CryptoUnets}

\begin{thebibliography}{}

\bibitem[Brakerski et~al., 2014]{brakerski2014leveled}
Brakerski, Z., Gentry, C., and Vaikuntanathan, V. (2014).
\newblock (leveled) fully homomorphic encryption without bootstrapping.
\newblock {\em ACM Transactions on Computation Theory (TOCT)}, 6(3):1--36.

\bibitem[Cheon et~al., 2017]{cheon2017homomorphic}
Cheon, J.~H., Kim, A., Kim, M., and Song, Y. (2017).
\newblock Homomorphic encryption for arithmetic of approximate numbers.
\newblock In {\em International Conference on the Theory and Application of
  Cryptology and Information Security}, pages 409--437. Springer.

\bibitem[Cheon et~al., 2022]{cheon2022efficient}
Cheon, J.~H., Kim, W., and Park, J.~H. (2022).
\newblock Efficient homomorphic evaluation on large intervals.
\newblock {\em IEEE Transactions on Information Forensics and Security},
  17:2553--2568.

\bibitem[Chiang, 2022]{chiang2022novel}
Chiang, J. (2022).
\newblock Volley revolver: A novel matrix-encoding method for
  privacy-preserving neural networks (inference).
\newblock {\em arXiv preprint arXiv:2201.12577}.

\bibitem[Chiang, 2023]{chiang2023privacy3layer}
Chiang, J. (2023).
\newblock Privacy-preserving cnn training with transfer learning: Multiclass
  logistic regression.
\newblock {\em arXiv preprint arXiv:2304.03807}.

\bibitem[Gentry, 2009]{gentry2009fully}
Gentry, C. (2009).
\newblock Fully homomorphic encryption using ideal lattices.
\newblock In {\em Proceedings of the forty-first annual ACM symposium on Theory
  of computing}, pages 169--178.

\bibitem[Han et~al., 2019]{han2018efficient}
Han, K., Hong, S., Cheon, J.~H., and Park, D. (2019).
\newblock Logistic regression on homomorphic encrypted data at scale.
\newblock In {\em Proceedings of the AAAI Conference on Artificial
  Intelligence}, volume~33, pages 9466--9471.

\bibitem[Jiang et~al., 2018]{kim2018matrix}
Jiang, X., Kim, M., Lauter, K., and Song, Y. (2018).
\newblock Secure outsourced matrix computation and application to neural
  networks.
\newblock In {\em Proceedings of the 2018 ACM SIGSAC Conference on Computer and
  Communications Security}, pages 1209--1222.

\bibitem[Kim et~al., 2018]{IDASH2018Andrey}
Kim, A., Song, Y., Kim, M., Lee, K., and Cheon, J.~H. (2018).
\newblock Logistic regression model training based on the approximate
  homomorphic encryption.
\newblock {\em BMC medical genomics}, 11(4):83.

\bibitem[Long et~al., 2015]{long2014fully}
Long, J., Shelhamer, E., and Darrell, T. (2015).
\newblock Fully convolutional networks for semantic segmentation.
\newblock In {\em Proceedings of the IEEE conference on computer vision and
  pattern recognition}, pages 3431--3440.

\bibitem[Nandakumar et~al., 2019]{nandakumar2019towards}
Nandakumar, K., Ratha, N., Pankanti, S., and Halevi, S. (2019).
\newblock Towards deep neural network training on encrypted data.
\newblock In {\em Proceedings of the IEEE/CVF conference on computer vision and
  pattern recognition workshops}, pages 0--0.

\bibitem[Ronneberger et~al., 2015]{ronneberger2015u}
Ronneberger, O., Fischer, P., and Brox, T. (2015).
\newblock U-net: Convolutional networks for biomedical image segmentation.
\newblock In {\em Medical image computing and computer-assisted
  intervention--MICCAI 2015: 18th international conference, Munich, Germany,
  October 5-9, 2015, proceedings, part III 18}, pages 234--241. Springer.

\bibitem[Smart and Vercauteren, 2011]{SmartandVercauteren_SIMD}
Smart, N. and Vercauteren, F. (2011).
\newblock Fully homomorphic simd operations.
\newblock Cryptology ePrint Archive, Report 2011/133.
\newblock \url{https://ia.cr/2011/133}.

\end{thebibliography}
\bibliographystyle{apalike}  

\end{document}